\documentclass[aps,prd,twocolumn,superscriptaddress,floatfix,amsmath,amssymb]{revtex4-2}
\usepackage[english]{babel}
\usepackage{graphicx}
\usepackage{dcolumn}
\usepackage{bm}
\usepackage{verbatim}
\usepackage{mathrsfs}
\usepackage{natbib}
\usepackage{cancel}
\usepackage{color}
\usepackage[normalem]{ulem}

\newcommand{\integral}[3]{\int_{#2}^{#3} \text{d} #1}

\newcommand{\ket}[1]{\left| {#1} \right\rangle}

\newcommand{\ematriz}[3]{\left\langle {#1} \left|{#2}\right|{#3}\right\rangle}

\def\slashchar#1{\setbox0=\hbox{$#1$} 
\dimen0=\wd0 
\setbox1=\hbox{/} \dimen1=\wd1 
\ifdim\dimen0>\dimen1 
\rlap{\hbox to \dimen0{\hfil/\hfil}} 
#1 
\else 
\rlap{\hbox to \dimen1{\hfil$#1$\hfil}} 
/ 
\fi}

\newcommand{\BbbR}{\mathbb{R}}

\begin{document}
\title{A Little Excitement Across the Horizon}
\author{Keith K. Ng}
\email{lafcadio08@gmail.com, ORCID 0000-0002-6552-8549}
\affiliation{Department of Physics, Nanyang Technological University, 21 Nanyang Link, Singapore
637371}
\author{Chen Zhang}
\email{zhangvchen@gmail.com, ORCID 0000-0002-2650-6873}
\affiliation{Department of Physics and Astronomy, University of Waterloo, Waterloo, Ontario, N2L 3G1, Canada}
\affiliation{Department of Physics, University of Toronto,
60 St.~George St., Toronto, Ontario, Canada M5S 1A7}
\author{Jorma Louko}
\email{jorma.louko@nottingham.ac.uk, ORCID 0000-0001-8417-7679}
\affiliation{School of Mathematical Sciences, University of Nottingham, Nottingham NG7 2RD, UK}
\author{Robert B. Mann}
\email{rbmann@uwaterloo.ca, ORCID 0000-0002-5859-2227}
\affiliation{Department of Physics and Astronomy, University of Waterloo, Waterloo, Ontario, N2L 3G1, Canada}
\affiliation{Institute for Quantum Computing, University of Waterloo, Waterloo, Ontario, N2L 3G1, Canada}
\affiliation{Perimeter Institute for Theoretical Physics, 31 Caroline St N, Waterloo, Ontario, N2L 2Y5, Canada}

\begin{abstract}
We analyse numerically the transitions in an Unruh-DeWitt detector, coupled linearly to a massless scalar field, in radial infall in (3+1)-dimensional Schwarzschild spacetime. In the Hartle-Hawking and Unruh states, the transition probability attains a small local extremum near the horizon-crossing and is then moderately enhanced on approaching the singularity. In the Boulware state, the transition probability drops on approaching the horizon. The unexpected near-horizon extremum arises numerically from angular momentum superpositions, with a deeper physical explanation to be found. 
\end{abstract}
\date{October 2021; revised September 2022. 
Published in New J. Phys.\ \textbf{24}, 103018 (2022).}
\maketitle

\section{Introduction} 

Attempts to combine well-established tenets of black hole physics and well-established tenets of quantum physics -- that  black hole evaporation is unitary, well described by local semi-classical quantum field theory, and respects the equivalence principle -- have been a source of a number of paradoxes~\cite{Mathur:2009hf,Mann:2015luq}, culminating in recent years to the suggestion that the horizon will be replaced by a high curvature region, 
inducing a ``firewall'' that will destroy any observer attempting to fall into the black hole~\cite{Almheiri:2012rt}. 
In this firewall paradigm, quantum physics near the black hole horizon differs significantly from what ensues from a gravitational collapse in the conventional semiclassical picture without back-reaction \cite{Hawking:1975vcx}, well approximated at late times by the Unruh state~\cite{unruh1976}, or from placing a radiating black hole in thermal equilibrium with its environment, described by the Hartle-Hawking(-Israel) state \cite{Hartle:1976tp,Israel:1976ur}.

While there has been much effort and discussion in attempting to resolve this conundrum,  scant attention has been paid toward understanding in semi-classical terms the actual behaviour of objects crossing the horizon, without a firewall. This is of crucial importance, since the only way to determine if a black hole has a temperature  -- that is, the vacuum of spacetime outside it is in a thermal state -- is to consider the response of some object (a detector) placed in the vicinity of the black hole \cite{Hu:2012jr}. Such detectors are most straightforwardly modelled as two-level systems known as Unruh-DeWitt (UDW) detectors \cite{unruh1976,DeWitt1979}.

When both the detector and the state of the quantum field are stationary, a number of results about the detector's response are known \cite{Hodgkinson:2012mr,Hodgkinson:2014iua}; in particular, a static detector in the exterior of a Schwarzschild black hole responds thermally to the Hartle-Hawking(-Israel) state 
\cite{Hartle:1976tp,Israel:1976ur}. 
However, essentially nothing 
is known about the response of detectors that freely fall toward a black hole, with the  exception of a few studies in lower dimensions. 
One  such study considered a UDW detector falling radially toward spinless black hole in (2+1) dimensions \cite{Hodgkinson:2012mr}. In this case no regions in parameter space were found where the transition rate was thermal, in accord with the equivalence principle, but the detector was switched on and off in the region exterior to the black hole.
A study of the transition rate of a UDW detector infalling toward a $(1+1)$ dimensional Schwarzschild black hole provided numerical evidence of how its thermal properties are gradually lost during the infall \cite{Juarez-Aubry:2014jba,JuarezPhDThesis}, 
and a detector approaching a Cauchy horizon in $1+1$ dimensions was considered in~\cite{Juarez-Aubry:2021tae}. 
A recent study of the $(1+1)$ dimensional Schwarzschild black hole
\cite{Gallock-Yoshimura:2021yok}
considered the entanglement and mutual information
acquired by two freely falling detectors and found that such 
correlations can be harvested even when the detectors are causally disconnected by the event horizon. 

In this paper we investigate for the first time  the response of a UDW detector (linearly coupled to a scalar field) 
in radial free-fall 
toward a Schwarzschild black hole and across its horizon. We consider the Hartle-Hawking(-Israel) 
\cite{Hartle:1976tp,Israel:1976ur}, 
Unruh \cite{unruh1976}, and 
Boulware \cite{Boulware:1974dm}
states of the field, where in the last case the detector is switched off before horizon crossing due to the singularity of the Boulware state at the horizon. Far from the horizon, we find that the response approaches a Minkowski thermal state response for the Hartle-Hawking state and the Minkowski vacuum response for the other two states, 
as could be expected. Near the horizon, the response in the Hartle-Hawking and Unruh states attains a small local extremum after horizon crossing and is then moderately enhanced on approaching the black hole singularity; in the Boulware state, the response drops as the horizon is approached. Our  work involves pioneering numerical elements, and constitutes, to our knowledge, the first analysis of a time-and-space localised quantum system crossing the horizon of a (3+1)-dimensional black hole. 

We begin by recalling in Section 
\ref{sec:spacetime-and-detector}
background about the Schwarzschild spacetime and 
the radially infalling trajectory therein, 
and in Section \ref{sec:detector-prelim}
about the UDW detector. 
Section
\ref{sec:modes-and-vacua}
describes the Boulware, Hartle-Hawking and Unruh states in terms of the scalar field mode functions, 
and 
Section \ref{sec:response-function} gives 
the mode sum expressions for the detector's response function in each of these three states. 
Section \ref{sec:switching} describes how the detector's interaction with the field is switched on and off. 
The numerical results for the detector's response are presented in Section~\ref{sec:results}, 
and Section 
\ref{sec:peak-physical}
describes the role of the angular momentum sum in the small extrema near the horizon. 
Section \ref{sec:conclusions} gives concluding remarks. 
Technical supporting material is deferred to
four
appendices. 
We use throughout units in which $\hbar = c = k_{\text{B}} = 1$.

\section{Spacetime and Detector's Motion}
\label{sec:spacetime-and-detector}

The exterior Schwarzschild metric reads 
\begin{align}
ds^2 = -\left( 1-2M/r\right)dt^2 + \frac{dr^2}{1-2M/r}+ r^2 d\Omega^2, 
\label{eq:schw-metric}
\end{align}
where $M>0$ and $r>2M$, 
and $d\Omega^2= d\theta^2+\sin^2 \! \theta\, d\varphi^2$ is the metric on the round unit two-sphere. 
The Kruskal-Szekeres extension is obtained by introducing the new coordinates (we follow the conventions of~\cite{birrell-davies1982}) 
\begin{subequations}
\label{eq:kruskalcoords-def}
\begin{align}
\bar{u} &= - 4M \bigl(r/(2M)-1\bigr)^{1/2} \, e^{(r-t)/(4M)}, 
\\
\bar{v} &= 4M \bigl(r/(2M)-1\bigr)^{1/2} \, e^{(r+t)/(4M)}, 
\end{align}
\end{subequations}
where initially $\bar{u} < 0 < \bar{v}$, 
and then extending the coordinate range to $\bar{u} \bar{v} < (4M)^2$. 
The metric becomes 
\begin{align}
ds^2 = - (2M/r) \, e^{-r/(2M)}\, d\bar{u} \, d\bar{v} + r^2 d\Omega^2, 
\label{eq:kruskal-metric}
\end{align}
where $r$ is the unique solution to 
$- \bar{u} \bar{v} = (4M)^2 e^{r/(2M)} \bigl(r/(2M) - 1\bigr)$. 
The conformal diagram can be found for example in~\cite{birrell-davies1982}. 

We consider the part of the Kruskal-Szekeres extension where $\bar{v} >0$. 
This contains the original Schwarzschild exterior at $\bar{u} < 0$, 
the black hole interior at $0 < \bar{u} < (4M)^2 / \bar{v}$, 
and the black hole horizon at $\bar{u} =0$. 

We consider a detector on an infalling radial timelike 
geodesic that starts from infinity with asymptotically vanishing initial velocity. 
In the Kruskal coordinates, the trajectory reads
\begin{subequations}
\label{eq:radial-infall-kruskal}
\begin{align}
\bar{u}&=-4M \! \left( z-1\right) \exp\!\left( z+\tfrac{1}{2} z^2 +\tfrac{1}{3} z^3 \right),
\\
\bar{v}&=4M \! \left( z+1\right) \exp\!\left( - z+\tfrac{1}{2} z^2 -\tfrac{1}{3} z^3 \right),
\end{align}
\end{subequations}
where $z=(\tau/\tau_H)^{1/3}$,
$\tau \in (-\infty, 0)$ is the proper time, and $\tau_H = -\tfrac43 M$. 
The trajectory starts from  past timelike infinity in the asymptotic past, $\tau\to-\infty$, 
crosses the horizon at $\tau = \tau_H$, 
and hits the black hole singularity at $\tau\to0^-$. 
The exterior Schwarzschild part of the trajectory 
can be expressed in the exterior Schwarzschild coordinates via~\eqref{eq:kruskalcoords-def}, and similar 
expressions can be given for the black hole interior part of the trajectory in terms of interior Schwarzschild coordinates. 


\section{Detector Preliminaries}
\label{sec:detector-prelim} 

Our UDW detector \cite{unruh1976,DeWitt1979} 
is a two-level quantum system of spatially negligible size, moving through the spacetime on the timelike worldline $\mathsf{x}(\tau)$, parametrised by the proper time~$\tau$. The detector's Hilbert space is spanned by two orthonormal states, with the respective eigenenergies $0$ and $\Omega \in \BbbR\setminus\{0\}$, defined with respect to~$\tau$. For $\Omega>0$, the state with eigenenergy $0$ is the ground state and the state with eigenenergy $\Omega$ is the excited state; 
for $\Omega<0$, the roles of the states are reversed. 

The detector couples linearly to a quantised real scalar field~$\phi$. While simple, this model captures the essential features of light-matter interactions when angular momentum interchange is negligible~\cite{MartinMartinez:2012th,Alhambra:2013uja}. 

Working to first-order perturbation theory in the coupling constant, the probability of the detector to make a transition from the eigenenergy $0$ state to the eigenenergy $\Omega$ state
is a multiple of the \emph{response function\/} $F(\Omega)$, given by 
\begin{align}
F(\Omega)&= \int \text{d}\tau \, \text{d}\tau' 
\, \chi(\tau) \chi(\tau') 
\, e^{-i\Omega (\tau-\tau')} \, W(\tau,\tau'), 
\label{eq:responsefunction-formula}
\end{align}
where $W(\tau,\tau')=\ematriz{\Psi}{\phi(\mathsf{x}(\tau))\phi(\mathsf{x}(\tau'))}{\Psi}$ 
is the pullback of the scalar field's Wightman function to the detector's worldline, $\ket{\Psi}$ denotes the initial state of the field, and 
the real-valued switching function $\chi$ specifies how the interaction is turned on and off. Assuming that $\ket{\Psi}$ has a technical regularity property known as the Hadamard condition~\cite{Decanini:2005eg}, 
$W(\tau,\tau')$ is a well-defined distribution under mild assumptions about the detector's trajectory~\cite{hormander-book,Fewster:1999gj}, 
and then $F(\Omega)$ is well defined under mild assumptions about~$\chi$. 
As the factor relating $F(\Omega)$ to the probability depends only on the detector's internal structure, we (with minor abuse of terminology) refer to $F(\Omega)$ as the probability. 

When the state $\ket{\Psi}$ is a `vacuum' state associated with a complete orthonormal set of `positive frequency' field modes, $\{\Psi_\beta(\mathsf{x})\}$, 
the Wightman function has the mode sum expression 
$\ematriz{\Psi}{\phi(\mathsf{x})\phi(\mathsf{x}')}{\Psi}
= \sum_\beta \Psi_\beta(\mathsf{x}) \Psi_\beta^*(\mathsf{x'})$, 
and the response function can be written as the mode sum 
\begin{align}
F(\Omega)=\sum_\beta 
\left| \int \text{d}\tau \, \chi(\tau) \, e^{-i\Omega \tau} \, \Psi_\beta(\mathsf{x} (\tau))\right|^2 , 
\label{eq:resp-modesum-general}
\end{align}
which we shall employ in what follows. 
 
\section{Mode Functions and States}
\label{sec:modes-and-vacua}

We take the field $\phi$ to be a real massless Klein-Gordon field. 
In the Schwarzchild exterior, we may separate the field equation, 
$\Box \phi =0$, by the ansatz 
\begin{align}
u(t,r,\theta,\varphi) =\frac{1}{\sqrt{4\pi\omega}}
\phi_{\omega l}(r)Y_{lm}(\theta,\varphi)
e^{-i\omega t}, 
\end{align}
where $Y_{lm}$ are the scalar spherical harmonics and the positive separation parameter $\omega$ 
is the frequency with respect to Schwarzschild time. 
The differential equation for the radial function 
$\phi_{\omega l}$ may be conveniently written in terms of the 
tortoise radial coordinate $r^*=r+2M\ln[r/(2M)-1] \in (-\infty,\infty)$ and the  scaling $\phi_{\omega l}(r) = \rho_{\omega l}(r)/r$, 
giving the Schr\"odinger-like equation \cite{Leaver1986}
\begin{equation}
\label{torteq}
\frac{d^2}{dr^{*2}}\rho_{\omega l}(r) + \left[ \omega^2 - V_l(r) \right]\rho_{\omega l}(r) = 0 , 
\end{equation}
with the effective potential 
\begin{equation}
V_l(r)=\left( 1-\frac{2M}{r} \right) \left( \frac{l(l+1)}{r^2}+\frac{2M}{r^3} \right).
\end{equation}
As the effective potential approaches zero when $r^* \to \infty$ and $r^* \to -\infty$, 
corresponding respectively to $r\to\infty$ and $r\to 2M$, 
there exists a set of mode solutions for $\phi_{\omega l}$, denoted by 
$\Phi_{\omega l}^{in}$ and $\Phi_{\omega l}^{up}$, 
specified uniquely by the asymptotic behaviour 
\begin{subequations}
\label{eq:in-and-up-asymptotics}
\begin{align}
\label{innorm}
\Phi_{\omega l}^{in} &\sim \left\lbrace 
\begin{matrix}
B^{in}_{\omega l}r^{-1}e^{-i\omega r^*} & r \rightarrow 2M, \\
r^{-1}e^{-i\omega r^*} + A^{in}_{\omega l}r^{-1}e^{i\omega r^*} & r \rightarrow \infty,
\end{matrix}
\right.\\
\label{upnorm}
\Phi_{\omega l}^{up} &\sim \left\lbrace
\begin{matrix}
r^{-1}e^{i\omega r^*} + A^{up}_{\omega l}r^{-1}e^{-i\omega r^*} & r \rightarrow 2M, \\
B^{up}_{\omega l}r^{-1}e^{i\omega r^*} & r \rightarrow \infty . 
\end{matrix}
\right.
\end{align}
\end{subequations}

We consider three quantum states for the field. 
First, the Boulware (B) state~\cite{Boulware:1974dm}, 
which approaches the Minkowski vacuum at $r\to\infty$, but is singular on both the black hole horizon and on the white hole horizon. 
Second, the Hartle-Hawking (H) state~\cite{Hartle:1976tp,Israel:1976ur}, 
which represents a radiating black hole in thermal equilibrium with a heat bath at $r\to\infty$, and is regular over all of the Kruskal-Szekeres extension. 
Third, the Unruh (U) state~\cite{unruh1976}, 
which represents a Hawking-radiating black hole with no incoming radiation from the past null infinity, and which is singular at the white hole horizon but regular across the black hole horizon. 
In the B state we can follow the infalling detector only outside the horizon, but in the H and U states we can follow the detector as it crosses the black hole horizon and into the black hole interior. 

The mode functions defining the B, H and U states in the exterior region 
can be written in terms of $\Phi_{\omega l}^{in}$ and $\Phi_{\omega l}^{up}$ and their conjugates, as lucidly reviewed in~\cite{Casals:2012es}. 
The H and U modes can then be continued to the black hole interior in the lower half-plane in complexified~$\bar{u}$~\cite{unruh1976}.

$\phantom{xxx}$ 

\section{Response function}
\label{sec:response-function}

We are now ready to write out the mode sum expression 
\eqref{eq:resp-modesum-general} for the response function for each of the three states. 

Consider first a detector that only operates in the Schwarzschild exterior. 
We may choose without loss of generality the trajectory to be at $\theta=0$, which simplifies the spherical harmonics. 
The pull-back of the Wightman function to the detector's worldline then has the 
mode sum expressions \cite{HodgkinsonPhDThesis,Hodgkinson:2014iua} 
\begin{widetext}
\begin{subequations}
\begin{align}
W_B(\tau,\tau')&=\sum_{l=0}^{\infty}\integral{\omega}{0}{\infty} \, \frac{2l+1}{16\pi^2\omega} 
 \left(\tilde{\Phi}_{\omega l}^{up}(r)\tilde{\Phi}_{\omega l}^{up*}(r')\left(\frac{\bar{u}}{\bar{u}'}\right)^{i4M\omega} + \tilde{\Phi}_{\omega l}^{in}(r)\tilde{\Phi}_{\omega l}^{in*}(r')\left(\frac{\bar{v}}{\bar{v}'}\right)^{-i4M\omega}\right), 
\label{bwightman}\\
\label{hhwightman}
W_{H}(\tau,\tau')&=\sum_{l=0}^{\infty}\integral{\omega}{0}{\infty} \, \frac{2l+1}{16\pi^2\omega} \left(\frac{e^{4\pi M\omega}}{2\sinh (4\pi M\omega)}\tilde{\Phi}_{\omega l}^{up}(r)\tilde{\Phi}_{\omega l}^{up*}(r')\left(\frac{\bar{u}}{\bar{u}'}\right)^{i4M\omega} + \frac{e^{-4\pi M \omega}}{2\sinh (4\pi M\omega)}\tilde{\Phi}_{\omega l}^{up*}(r)\tilde{\Phi}_{\omega l}^{up}(r')\left(\frac{\bar{u}}{\bar{u}'}\right)^{-i4M \omega} \right. \nonumber \\
&\qquad\qquad +\left. \frac{e^{4\pi M\omega}}{2\sinh (4\pi M\omega)}\tilde{\Phi}_{\omega l}^{in}(r)\tilde{\Phi}_{\omega l}^{in*}(r')\left(\frac{\bar{v}}{\bar{v}'}\right)^{-i4M\omega} + \frac{e^{-4\pi M\omega}}{2\sinh (4\pi M\omega)}\tilde{\Phi}_{\omega l}^{in*}(r)\tilde{\Phi}_{\omega l}^{in}(r')\left(\frac{\bar{v}}{\bar{v}'}\right)^{i4M\omega}\right),\\
\label{uwightman}
W_U(\tau,\tau')&=\sum_{l=0}^{\infty}\integral{\omega}{0}{\infty} \, \frac{2l+1}{16\pi^2\omega} \left(\frac{e^{4\pi M\omega}}{2\sinh (4\pi M\omega)}\tilde{\Phi}_{\omega l}^{up}(r)\tilde{\Phi}_{\omega l}^{up*}(r')\left(\frac{\bar{u}}{\bar{u}'}\right)^{i4M\omega} + \frac{e^{-4\pi M\omega}}{2\sinh (4\pi M\omega)}\tilde{\Phi}_{\omega l}^{up*}(r)\tilde{\Phi}_{\omega l}^{up}(r')\left(\frac{\bar{u}}{\bar{u}'}\right)^{-i4M\omega}\right.\nonumber\\
& \qquad\qquad +\left.\tilde{\Phi}_{\omega l}^{in}(r)\tilde{\Phi}_{\omega l}^{in*}(r')\left(\frac{\bar{v}}{\bar{v}'}\right)^{-i4M\omega}\right),
\end{align}
\end{subequations}
where 
\begin{align}
\tilde{\Phi}_{\omega l}^{up}(r) & :=e^{-i \omega r^*}\Phi_{\omega l}^{up}(r) ,  \qquad 
\tilde{\Phi}_{\omega l}^{in}(r) := e^{+i \omega r^*}\Phi_{\omega l}^{in}(r) \, ,
\label{eq:Phitilde-definitions}
\end{align} 
and the subscripts B, H and U label the three states. 
Unprimed and primed coordinates are evaluated at $\tau$ and~$\tau'$ respectively. 
In each term, the factor depending on the 
unprimed coordinates comes from the `positive frequency' mode functions defining the corresponding state, and the factor depending on the primed coordinates comes from the complex conjugate `negative frequency' mode functions. 

From \eqref{eq:resp-modesum-general}, we now obtain 
\begin{equation}
F_B = 
F_B^{up}
+ F_B^{in},   \qquad 
F_H = 
F_H^{up} + F_H^{\widetilde{up}}
+ F_H^{in} + F_H^{\widetilde{in}},  \qquad 
F_U = 
F_H^{up} + F_H^{\widetilde{up}}
+ F_B^{in} , 
\label{eq:BHU-F}
\end{equation}
for the respective B, H, and U  states,
where 
\begin{subequations}
\label{eq:Fup-in-collected}
\begin{align}
F_B^{up}(\Omega)&=\sum_{l}\int_{0}^\infty d\omega \, \frac{2l+1}{16\pi^2 \omega}  \left| \integral{\tau}{}{}\,\chi(\tau)e^{-i\Omega \tau}\, \tilde{\Phi}^{up}_{\omega l}(r)\left(\frac{-\bar{u}}{4M}\right)^{i4M \omega}\right|^2, 
\label{eq:F-B-up}
\\
F_B^{in}(\Omega)&=\sum_{l}\int_{0}^\infty d\omega \, \frac{2l+1}{16\pi^2 \omega}  \left| \integral{\tau}{}{}\,\chi(\tau)e^{-i\Omega \tau} \, \tilde{\Phi}^{in}_{\omega l}(r)\left(\frac{\bar{v}}{4M}\right)^{-i4M \omega}\right|^2, 
\label{eq:F-B-in}
\\
F_H^{up}(\Omega)&=\sum_{l}\int_{0}^\infty d\omega \, \frac{2l+1}{16\pi^2 \omega} \times \frac{e^{4\pi M\omega}}{2\sinh (4\pi M\omega)}  \left| \integral{\tau}{}{}\,\chi(\tau)e^{-i\Omega \tau} \, \tilde{\Phi}^{up}_{\omega l}(r)\left(\frac{-\bar{u}}{4M}\right)^{i 4M \omega}\right|^2, 
\label{eq:F-H-up-normal}
\\
F_H^{\widetilde{up}}(\Omega)&=\sum_{l}\int_{0}^\infty d\omega \, \frac{2l+1}{16\pi^2 \omega} \times \frac{e^{-4\pi M\omega}}{2\sinh (4\pi M\omega)}  \left| \integral{\tau}{}{}\,\chi(\tau)e^{-i\Omega \tau}\,\tilde{\Phi}^{up*}_{\omega l}(r)\left(\frac{- \bar{u}}{4M}\right)^{- i 4M \omega}\right|^2, 
\label{eq:F-H-up-tilded}
\\
F_H^{in}(\Omega)&=\sum_{l}\int_{0}^\infty d\omega \, \frac{2l+1}{16\pi^2 \omega} \times \frac{e^{4\pi M\omega}}{2\sinh (4\pi M\omega)}  \left| \integral{\tau}{}{}\,\chi(\tau)e^{-i\Omega \tau}\,\tilde{\Phi}^{in}_{\omega l}(r)\left(\frac{\bar{v}}{4M}\right)^{-i 4M \omega}\right|^2, 
\label{eq:F-H-in-normal}
\\
F_H^{\widetilde{in}}(\Omega)&=\sum_{l}\int_{0}^\infty d\omega \, \frac{2l+1}{16\pi^2 \omega} \times \frac{e^{-4\pi M\omega}}{2\sinh (4\pi M\omega)}  \left| \integral{\tau}{}{}\,\chi(\tau)e^{-i\Omega \tau}\,\tilde{\Phi}^{in*}_{\omega l}(r)\left(\frac{\bar{v}}{4M}\right)^{i 4M \omega}\right|^2. 
\label{eq:F-H-in-tilded}
\end{align}
\end{subequations}

Consider then a detector that continues to operate behind the horizon. As the B state is singular on the horizon, we consider only the H and U states. 
As shown in Appendix~\ref{app:continuation}, 
following~\cite{HodgkinsonPhDThesis}, 
and as discussed in the context of vacuum polarisation in~\cite{Lanir:2018rap}, 
expressions  
valid both in the exterior and interior for $F_H^{up}$ and $F_H^{\widetilde{up}}$  are 
\begin{subequations}
\label{eq:F-H-up-continued}
\begin{align}
F_H^{up}(\Omega)&=\sum_{l}\int_{0}^\infty d\omega \, \frac{2l+1}{16\pi^2 \omega} \times \frac{e^{4\pi M\omega}}{2\sinh (4\pi M\omega)} 
\times
\left| \integral{\tau}{}{}\,\chi(\tau)e^{-i\Omega \tau}
\left\{ \Theta(-\bar{u})\tilde{\Phi}^{up}_{\omega l}(r)\left(\frac{-\bar{u}}{4M}\right)^{i 4M \omega} 
\right. \right. 
\notag
\\
& \hspace{6ex}
\left.\left.
+ \Theta(\bar{u}) \left[ \frac{A^{up} \tilde{\Phi}^{in}_{\omega l}(r)}{B^{in}} \left(\frac{\bar{v}}{4M}\right)^{-i 4M \omega}
+ \frac{e^{-4\pi M\omega}\tilde{\Phi}^{in*}_{\omega l}(r)}{B^{in*}} \left(\frac{\bar{u}}{4M}\right)^{i 4M \omega} \right] 
\right\} \right|^2, 
\\
F_H^{\widetilde{up}}(\Omega)&=\sum_{l}\int_{0}^\infty d\omega \, \frac{2l+1}{16\pi^2 \omega} \times \frac{e^{-4\pi M\omega}}{2\sinh (4\pi M\omega)} 
\times
\left| \integral{\tau}{}{}\,\chi(\tau)e^{-i\Omega \tau}
\left\{ \Theta(-\bar{u})\tilde{\Phi}^{up*}_{\omega l}(r)\left(\frac{-\bar{u}}{4M}\right)^{-i 4M \omega} 
\right. \right. 
\notag
\\
& \hspace{6ex}
\left.\left.
+ \Theta(\bar{u}) \left[ \frac{A^{up*} \tilde{\Phi}^{in*}_{\omega l}(r)}{B^{in*}} \left(\frac{\bar{v}}{4M}\right)^{i 4M \omega}
+ \frac{e^{4\pi M\omega}\tilde{\Phi}^{in}_{\omega l}(r)}{B^{in}} \left(\frac{\bar{u}}{4M}\right)^{-i 4M \omega} \right] 
\right\} \right|^2, 
\end{align}
\end{subequations}
\end{widetext}
where $\Theta$ is the Heaviside theta-function.
The responses for the H and U states are hence given by $F_H$ and $F_U$ in~\eqref{eq:BHU-F}, 
using \eqref{eq:F-B-in}, \eqref{eq:F-H-in-normal}, 
\eqref{eq:F-H-in-tilded}
and~\eqref{eq:F-H-up-continued}.

\section{Switching}
\label{sec:switching}

We switch the interaction on and off with 
\begin{align}
\chi(\tau)&=\begin{cases}
{\displaystyle \cos^4 \! \left(\frac{\pi(\tau-\tau_{mid})}{2\Delta}\right)} &
\text{for} \ \ |\tau-\tau_{mid}| < \Delta, \\
0 & \textrm{otherwise}, 
\end{cases}
\label{eq:switching-family}
\end{align}
where the positive parameter $\Delta$ is half the interaction duration and the parameter $\tau_{mid}$ is the midpoint of the interaction interval. 
$\chi$~is a close approximation to Gaussian switching over its duration~\cite{Cong:2020crf}, and it is smooth everywhere except at the switch-on and switch-off moments, where it is still~$C^3$, which is sufficiently regular to render the response finite. 

Note that $\chi$ has peak value~$1$, regardless of the total duration, with the convenient consequence that the response function is dimensionless, but with the price that the overall magnitudes of the response function for differing values of $\Delta$ are not directly comparable. 
Factoring out an `overall duration' effect could be accomplished by including a normalisation factor proportional to $\Delta^{-1/2}$ \cite{Martin-Martinez:2014qda,Fewster:2016ewy}. 

\section{Results}
\label{sec:results} 

We present here graphs of the response as the detector falls close to and in the black hole. 
The numerical implementation is described in Appendices 
\ref{app:radial-numerical} and~\ref{app:convergence}.

\begin{figure}
    \centering
    \includegraphics[width=0.99\columnwidth]{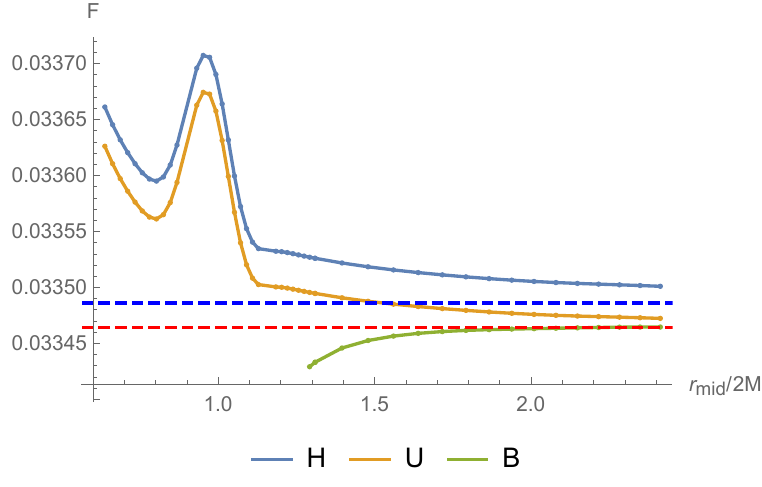}
    \caption{Response for the three states with  $\Omega=5/(2M)$ and $\Delta=0.3\times (2M)$, as a function of $r_{mid} = r(\tau_{mid})$. Note that the response for the Boulware state is defined only when the full interval in which the detector operates is outside the horizon. The lower (red) and upper (blue) dashed horizontal lines indicate respectively the flat space values $F_{vac}$ and $F_{\beta}$, described in the text.}
    \label{fig:vacuumgraph}
\end{figure}

First, we plot in Figure \ref{fig:vacuumgraph} the response function 
as a function of $r_{mid}=r(\tau_{mid}),$ for detector frequency $\Omega = 5/(2M)$, and switching time $\Delta = 0.3\times 2M$, with respect to all three states.  Far from the horizon, the response function approaches that of a detector in flat space: for the B and U states, the asymptotic value is the flat space vacuum value~$F_{vac}\approx 0.0334638$, and for the H state, the asymptotic value is the flat space thermal state value $F_{\beta}\approx 0.0334861$ in the Hawking temperature $1/(8\pi M)$ (these values are computed in Appendix~\ref{app:minkowski-comparison}). 
This is as expected: note that the difference between the H and U states arises because the H state is in equilibrium   with a thermal bath that extends to infinity, whereas in the U state the black hole is emitting net radiation whose intensity falls off far from the hole. We likewise see that the response in the B state falls off as the horizon is approached. 
The maximum departure from these asymptotic values is of the order of parts per thousand.  This shows that the major part of the response in our parameter range is due to the finiteness of the time interval during which the detector operates. 

\begin{figure}
    \centering
    \includegraphics[width=0.99\columnwidth]{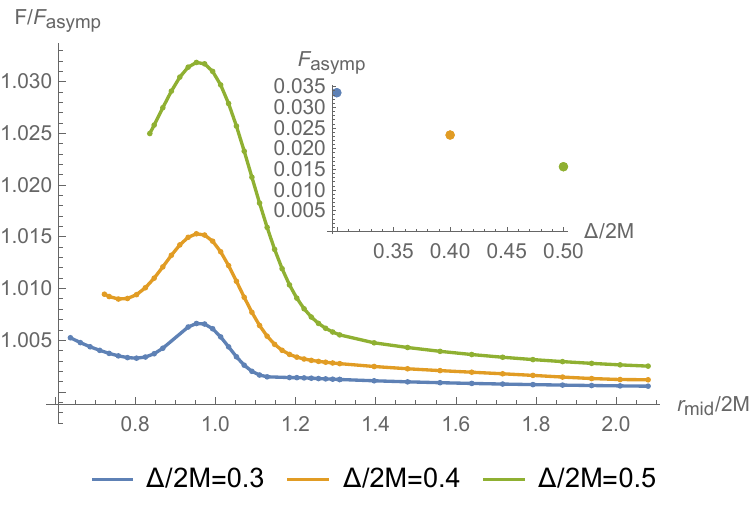}
    \caption{The normalized response $F/F_{asymp}$, where $F_{asymp}$ (inset) is the response 
    as $r_{mid}\to\infty$ (obtained from the analytic result in Appendix~\ref{app:minkowski-comparison}), for selected values of the switching duration parameter~$\Delta$, with $\Omega=5/(2M)$, in the Hartle-Hawking state. Lines end on the left when detector switchoff is at the singularity.}
    \label{fig:deltagraph}
\end{figure}

\begin{figure}
    \centering
    \includegraphics[width=0.99\columnwidth]{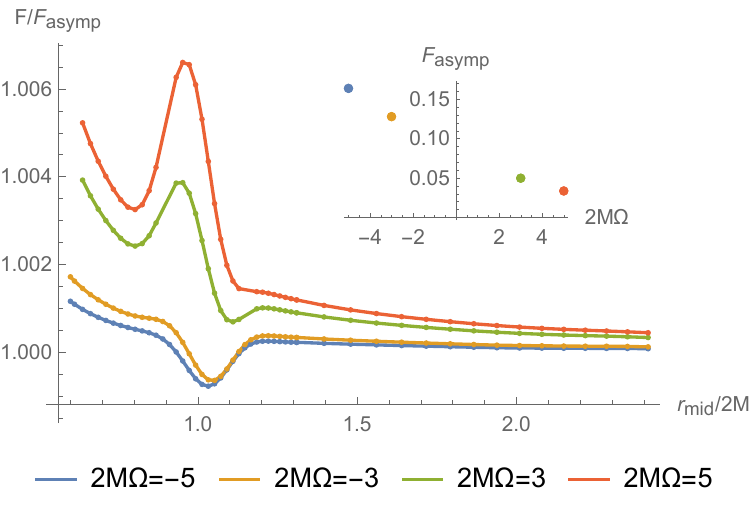}
    \caption{ The normalized response $F/F_{asymp}$, where $F_{asymp}$ (inset) is the response as $r_{mid}\to\infty$ (obtained from the analytic result in Appendix~\ref{app:minkowski-comparison}) for selected values of the energy gap $\Omega$, with $\Delta = 0.3\times 2M$, in the Hartle-Hawking state.}
    \label{fig:omegagraph}
\end{figure}

We also observe (for H and U) a small but discernible peak in the response as the detector crosses the  horizon. Although this region 
involves some numerical subtlety in the matching in \eqref{eq:F-H-up-continued} across the horizon, we have checked 
(see Appendix~\ref{app:convergence}) 
that the peak is robust on varying the numerical parameters. 
We shall address the origins of this peak in Section~\ref{sec:peak-physical}. 

Enhancement of radiation seen by an infalling observer approaching the horizon has been noted previously using an effective temperature function~\cite{Barbado:2011dx}, 
and a near-horizon discussion based on the equivalence principle has been given in~\cite{Ben-Benjamin:2019opz}. 
The relationship with our approach remains an interesting open question. 

Next, we plot in Figure \ref{fig:deltagraph} the response for selected values of $\Delta$ in the H state. 
We again observe a small but discernible  peak in the response that is $\Delta$-dependent.
A curious tradeoff becomes visible: 
if $\Delta$ is too small, the finer features of the states are overwhelmed by switching effects \cite{Fewster:2016ewy}; but
$\Delta$ cannot be too large since the detector cannot remain `switched on' at the $r=0$ singularity.
 These facts conspire to limit detector sensitivity  to the  state of the field. (In fact, since the B state is  defined only outside the horizon,  sensitivity in that case is reduced even further.)

Finally, in Figure \ref{fig:omegagraph} we consider the effect of varying the detector's energy gap $\Omega$, recalling that $\Omega>0$ corresponds to excitations and $\Omega<0$ to de-excitations. We specialise to the H state. 
The response as a function of $r_{mid}$ exhibits local extrema near and past $r_{mid} = 2M$, but the detailed loci and character of these extrema depend on~$\Omega$. When the detector falls deeper in the black hole, the response exhibits a modest overall growth (also evident in Figure~\ref{fig:vacuumgraph}). The magnitude of the $r_{mid}$-dependent variation is however small, of the order of parts per thousand, in the parameter range that we have explored.

\section{Origins of the peak}
\label{sec:peak-physical} 

In this section we present evidence that the small local extremum in the detector's response near the horizon is a cumulative effect from superposition of low and high angular momentum field modes, 
having no counterpart in the corresponding $(1+1)$-dimensional infall, 
where the field only has the zero angular momentum sector. 
This suggests that the angular momentum of the field is an essential contributor to the extremum.

\subsection{Detector infall in $1+1$ dimensions}
\label{sec:twodim-comparison} 

First, for comparison, we consider the corresponding detector infall in the $(1+1)$-dimensional Schwarzschild spacetime, whose metric is given by the first two terms in~\eqref{eq:schw-metric}. The field now has only the zero angular momentum sector. 

\begin{figure}
    \centering
    \includegraphics[width=0.9\linewidth]{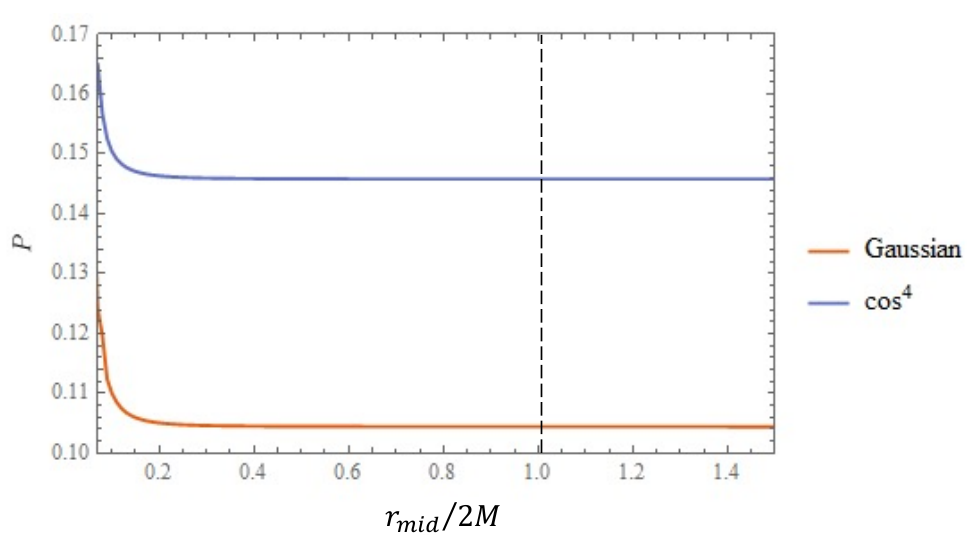}
    \caption{Response of a detector falling through the horizon (vertical black line) of the $(1+1)$-dimensional 
Schwarzschild spacetime in the Unruh state  for
a Gaussian switching function $e^{-\frac{(\tau-\tau_{mid})^2}{\Delta^2}}$ (orange, lower curve) and a $\cos^4(\frac{\pi(\tau-\tau_{mid})}{2\Delta})$ switching function (blue, higher curve). The energy gap is $\Omega\Delta = 1$. $r_{mid}$ denotes the location of the peak of the switching.}
    \label{2Dresponse}
\end{figure}

The Wightman functions of a massless scalar in the B, U and H states are elementary functions, and 
the infrared ambiguity in these Wightman functions, typical of a massless field in $1+1$ dimensions, drops out when we consider a detector that couples linearly to the proper time derivative of the field, rather than to the value of the field~\cite{Juarez-Aubry:2014jba}. 
Note also that the short-distance properties of a derivative-coupled detector in $1+1$ dimensions are similar to those of a non-derivative detector in $3+1$ dimensions~\cite{Gallock-Yoshimura:2021yok}.

The numerical codes to analyse this situation can be readily extracted from those that were used in \cite{Gallock-Yoshimura:2021yok}, 
by a team including one of the present authors, 
to analyse detector-detector entanglement during the infall. 
We exhibit the response for the U state in Figure~\ref{2Dresponse}, for both the cosine switching function \eqref{eq:switching-family}
and the corresponding Gaussian switching function used in~\cite{Gallock-Yoshimura:2021yok}.  
The behaviour is monotonic across the horizon, exhibiting no local extremum.

\subsection{High and low angular momenta in $3+1$ dimensions}
\label{sec:ell-summation} 

Second, returning to $3+1$ dimensions, we consider the contributions to the response from low and high angular momenta. 

\begin{figure}
    \centering
    \includegraphics[width=0.95\linewidth]{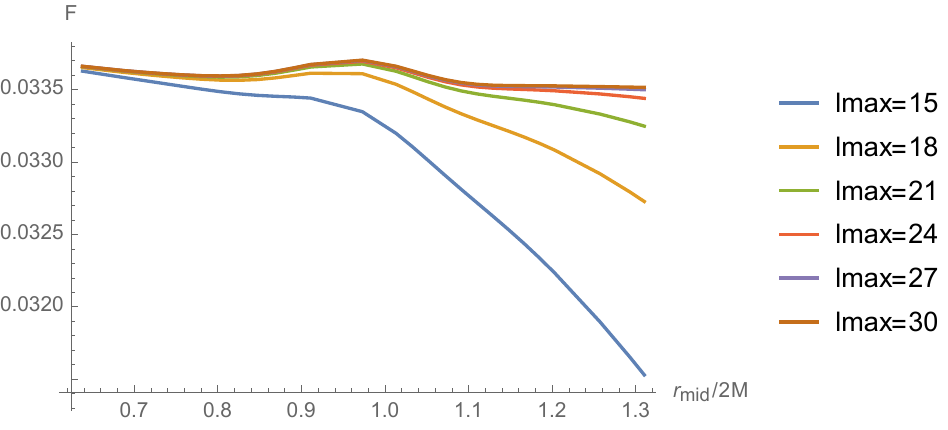}
    \caption{Detector response contributions in the Hartle-Hawking state with high $l$ cutoff $l_{max}$ as a function of $r_{mid}/2M.$}
    \label{fig:lpartial}
\end{figure}

The detector's response involves a sum over the angular momentum quantum number~$l$, 
and the graphs in 
Figures \ref{fig:vacuumgraph}--\ref{fig:omegagraph}
were obtained by increasing the high $l$ cutoff $l_{max}$ until the results stabilised. 
To examine the effect of the high $l$ modes, 
we plot in Figure \ref{fig:lpartial} partial sums, for 
selected values of~$l_{max}$. 
The horizontal axis is $r_{mid}/(2M)$, 
as in Figures \ref{fig:vacuumgraph}--\ref{fig:omegagraph}, and the state is the H state. 

For low values of~$l_{max}$, Figure \ref{fig:lpartial} shows that the small local extremum near the horizon is no longer present. 
The bump only appears for sufficiently high~$l_{max}$. 
Angular momentum hence appears essential for the bump. 

\begin{figure}
    \centering
    \includegraphics[width=0.93\linewidth]{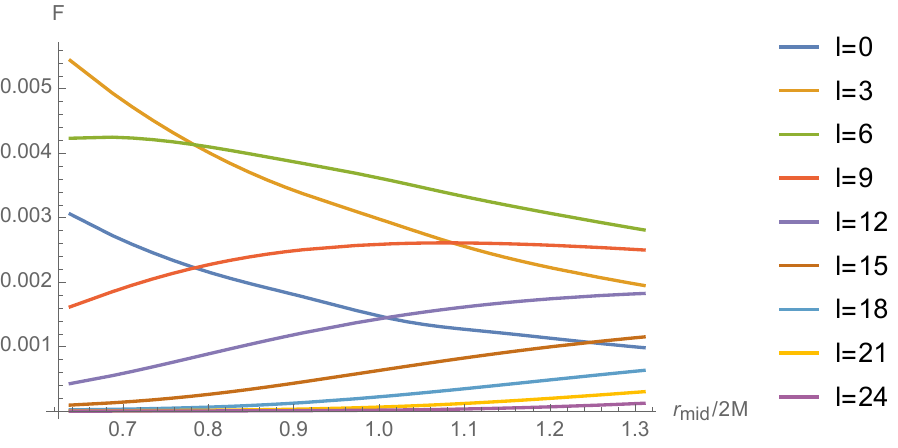}
    \caption{Detector response contribution in the Hartle-Hawking state from selected individual values of $l$ as a function of $r_{mid}/2M.$}
    \label{fig:lindividual}
\end{figure}

To examine the individual contributions of different angular momentum modes, we plot in Figure \ref{fig:lindividual} selected individual $l$ terms in the detector's response, for the H state. 

The low-$l$ modes (including the S-wave mode, $l=0$) contribute the most to the response closer to the singularity, while the contribution of the higher $l$ modes is larger far from the black hole. The bump near the horizon is the superposition of these two effects. We emphasise that the  contribution of the non-S-wave modes is of significant magnitude, and this contribution must be included to evaluate the response accurately.

\section{Conclusions}
\label{sec:conclusions} 

We have evaluated numerically, for the first time, the response of a time-and-space localised quantum detector 
falling through the horizon of a Schwarzschild black hole. 
When the ambient quantum field is in the 
Hartle-Hawking state or in the Unruh state, 
the detector's response remains regular across the horizon, 
as indeed had to happen by the regularity properties of these states. 
However, the response displays an unexpected local extremum near the horizon-crossing, at a few parts per thousand in the parameter regions we have explored.
Numerical evidence indicates that this extremum comes from the superposition of contributions from low and high angular momentum modes of the field, 
and it has no counterpart
for the analogous detector infall in $(1+1)$ dimensions. 
A~deeper physical explanation of this extremum remains an intriguing topic for further study.
For example, is a similar feature present in all spacetime dimensions higher than two, and for quantum fields of any spin? Also, might the detailed form of the switching function have a role in this feature? 

Our numerical parameter range covered the detector's locus inwards from area-radius slightly above~$4M$, 
the detector's interaction duration from $1.2M$ to~$2M$, 
and the detector's energy gap from $3/(2M)$ to $5/(2M)$, 
for excitations and de-excitations. 
These ranges are of order unity, in the units of~$M$, and they focus on a regime where varying the parameters did reveal characteristic features, yet kept the numerical demands manageable. The energy gaps probed are near the characteristic energy in the Hawking radiation; going to much higher or much lower values of the gap would increase the numerical demands. The maximum value of the area-radius is large enough to indicate numerical convergence towards the analytically-known asymptotic values at infinity, as discussed in Section \ref{sec:results} and Appendix~\ref{app:minkowski-comparison}; 
going to higher values of the area-radius would increase the numerical demands, but would not be expected to uncover qualitatively new features. 
The detector's interaction duration is small enough to reveal time-dependent features near and past the horizon; going to smaller durations would improve the time resolution, but it would also increase the pure switching contribution to the detector's response, and require higher numerical accuracy to distinguish the contributions due to the motion and to the field's quantum state, which contributions are the interesting ones. We recall that with the current parameters, the contributions due to motion and the field's quantum state are less than one part per thousand far from the hole, raising to a few parts per thousand near and in the hole. 

To rephrase, our numerical parameter range was chosen for demonstrating a fundamental prediction of the theory, rather than as a proposal for a practical observation or experiment. For example, for a solar mass black hole, $M\sim 10^3$m, our detector has total operating time $\sim 10^{-5}$s and energy gap $\sim 10^{-10}$eV, 
whereas for a galactic centre black hole, 
$M\sim 10^{10}$m, the total operating time is $\sim 100$s and the energy gap is $\sim 10^{-17}$eV. 
These values seem difficult to relate to astrophysical observations of black holes. 

One way to generalise our analysis would be to vary the geodesic trajectory of the detector. For example, the limitations due to the detector's finite interaction duration 
could be ameliorated by considering trajectories that spend more proper time in the near-horizon region. Alternatively, the region between the closed photon orbit at $r=3M$ and the innermost stable circular timelike orbit at $r=6M$ can be explored by nonradial high-eccentricity geodesics that come from infinity and return there. Perhaps Alice the astronaut can gain significant information by just observing the dance of quantum matter around the hole, without having to dive into the hole herself. 

Our methods are applicable to other quantum states and to other spacetimes. 
For instance, the methods can be used to evaluate the response of a detector to quantum states defined on proposed alternatives to black holes, such as those discussed in~\cite{Holdom:2020uhf}.
This may provide a quantum tool for distinguishing a black hole from its alternatives, and by extension a tool for testing quantum extensions to general relativity, offering the prospect of  taking one more step towards a unified theory.

\section*{Acknowledgments}

We thank Luis Barbado and anonymous referees for helpful comments. 
KN~thanks the School of Mathematical Sciences at the University of Nottingham for hospitality in the early stages of the work. 
This work was supported in part by the Natural Science and Engineering Research Council of Canada and by Asian Office of Aerospace Research and Development Grant No.\ FA2386-19-1-4077.
The work of JL was supported by United Kingdom Research and Innovation 
Science and Technology Facilities Council 
[grant numbers ST/P000703/1, ST/S002227/1]. 
For the purpose of open access, the authors have applied a CC BY public copyright licence to any Author Accepted Manuscript version arising.

\section*{Data availability}

The data that support the findings of this study are available upon reasonable request from the authors.

\section*{Conflict of interest}

The authors have no conflicts to disclose.

\appendix

\section{Continuation across the horizon}
\label{app:continuation} 

In this appendix we verify the analytic continuation of $F_H^{up}$ and $F_H^{\widetilde{up}}$ from the exterior formulas \eqref{eq:F-H-up-normal} and \eqref{eq:F-H-up-tilded} 
to the formulas \eqref{eq:F-H-up-continued} that are valid also inside the horizon, for the evaluation of the response in the H and U states. 

First, note that each of the expressions in \eqref{eq:Fup-in-collected} is written so that the quantity within the absolute values comes from the `positive frequency' 
mode functions characterising the respective state. 

Next, recall that the `positive frequency' mode functions defining the H and U  states have the property that if they are singular on the horizon, where $\bar{u}$ changes sign, they are continued across the horizon in the lower half of the complex $\bar{u}$ plane 
\cite{unruh1976,birrell-davies1982,Casals:2012es}. 
As $\bar{v}>0$ in the spacetime regions in which we are working, and as $\tilde{\Phi}^{in}_{\omega l}$ \eqref{eq:Phitilde-definitions}
is smooth across the black hole horizon, the expressions for $F_B^{in}$, $F_H^{in}$ and $F_H^{\widetilde{in}}$ in \eqref{eq:Fup-in-collected}
remain well defined and suitable for numerical evaluation at and behind the horizon. 
By contrast, the expressions for $F_H^{up}$ and $F_H^{\widetilde{up}}$ in \eqref{eq:Fup-in-collected} need to be given an appropriate analytic continuation, as described in~\cite{HodgkinsonPhDThesis}, 
and as we now explain. 
See 
\cite{Lanir:2018rap} 
for a discussion of this continuation in the context of vacuum polarisation.

Consider the exterior formulas \eqref{eq:F-H-up-normal} and \eqref{eq:F-H-up-tilded} for $F_H^{up}$ and~$F_H^{\widetilde{up}}$. 
Under each integral, we write (where our convention for $B^{in}$ differs from 
\cite{HodgkinsonPhDThesis})
\begin{widetext}
\begin{subequations}
\label{eq:tildePhi-continuationready}
\begin{align}
\tilde{\Phi}^{up}_{\omega l}(r)\left(\frac{-\bar{u}}{4M}\right)^{i 4M \omega}
& = 
\frac{A^{up} \tilde{\Phi}^{in}_{\omega l}(r)}{B^{in}} \left(\frac{\bar{v}}{4M}\right)^{-i 4M \omega}
+ \frac{\tilde{\Phi}^{in*}_{\omega l}(r)}{B^{in*}} \left(\frac{-\bar{u}}{4M}\right)^{i 4M \omega}, 
\\
\tilde{\Phi}^{up*}_{\omega l}(r)\left(\frac{-\bar{u}}{4M}\right)^{-i 4M \omega}
& = 
\frac{A^{up*} \tilde{\Phi}^{in*}_{\omega l}(r)}{B^{in*}} \left(\frac{\bar{v}}{4M}\right)^{i 4M \omega}
+ \frac{\tilde{\Phi}^{in}_{\omega l}(r)}{B^{in}} \left(\frac{-\bar{u}}{4M}\right)^{-i 4M \omega}, 
\end{align}
\end{subequations}
\end{widetext}
using
\eqref{eq:in-and-up-asymptotics}
and~\eqref{eq:Phitilde-definitions}, 
and recalling that $\bar{u}<0$. 
Continuing now to positive $\bar{u}$ in the lower half of the complex $\bar{u}$ plane, 
the respective first terms on the right-hand sides in \eqref{eq:tildePhi-continuationready} 
remain well defined as written, 
whereas the second terms respectively continue to 
\begin{subequations}
\begin{align}
&\frac{e^{-4\pi M\omega}\tilde{\Phi}^{in*}_{\omega l}(r)}{B^{in*}} \left(\frac{\bar{u}}{4M}\right)^{i 4M \omega} , 
\\
&\frac{e^{4\pi M\omega}\tilde{\Phi}^{in}_{\omega l}(r)}{B^{in}} \left(\frac{\bar{u}}{4M}\right)^{-i 4M \omega} . 
\end{align}
\end{subequations}
This gives the expressions \eqref{eq:F-H-up-continued} for 
$F_H^{up}$ and~$F_H^{\widetilde{up}}$.

\section{Numerical integration of the radial equation}
\label{app:radial-numerical} 

In this appendix we describe the numerical integration of the radial equation~\eqref{torteq}. 

The form of \eqref{torteq} suggests an obvious strategy: take initial conditions in the ``asymptotic free'' regime, then numerically integrate the differential equation. However, things are somewhat complicated by a few issues. The most obvious of these is that the modes are defined and normalized on `opposite sides'; for instance, while the in mode is defined by pure ingoing behaviour near the horizon, it is normalized in the other asymptotic regime, far from the horizon. In this section, we use $\Phi$ to refer to modes normalized as in~\eqref{eq:in-and-up-asymptotics}, 
and $\phi$ for an alternate normalization, which will be defined shortly.

When we enter initial values into the differential equation, we must use the asymptotic regime where the mode is expressible as a single exponential; more specifically, we use an initial value such that
\begin{subequations}
\begin{align}
\phi_{\omega l}^{in} &\sim r^{-1} e^{-i\omega r^*}, & r \rightarrow 2M , 
\\
\phi_{\omega l}^{up} &\sim r^{-1} e^{i\omega r^*}, & r \rightarrow \infty . 
\end{align}
\end{subequations}
However, this normalization is clearly at odds with the normalization required for $\Phi$.

In order to resolve this problem, we can introduce two further sets of solutions: the downward and outward modes, which we define by
\begin{subequations}
\begin{align}
\phi_{\omega l}^{out} &\sim r^{-1} e^{i\omega r^*}, & r \rightarrow 2M, 
\\
\phi_{\omega l}^{down} &\sim r^{-1} e^{-i\omega r^*}, & r \rightarrow \infty . 
\end{align}
\end{subequations}
Normalization is then a matter of expressing each physical mode in terms of a superposition of two other modes; this allows us to deduce the correct factor by which we must multiply $\phi$ to yield~$\Phi$. 
In essence, we are solving for the transmission and reflection coefficients 
in~\eqref{eq:in-and-up-asymptotics}. 

While we could, in principle, simply take the boundary conditions at face value for some suitably asymptotic finite values of $r^*$, there are more precise estimates of the modes in asymptotic regimes available. Following Leaver~\cite{Leaver1986},  
we can make estimates for the up and down modes far from the horizon as follows:
\begin{widetext}
\begin{subequations}
\begin{align}
\phi_{\omega l}^{up} &\sim \frac{1}{\sqrt{r(r- {2M})}} H^+_{\nu_a}(-2M\omega,\omega r) 
\exp \! \left[ i \left(-2M\omega \log (4M\omega) +  \tfrac{\pi}{2}\nu_a - \sigma_{\nu_a}(-{2M}\omega)\right) \right], \\
\phi_{\omega l}^{down} &\sim \frac{1}{\sqrt{r(r-{2M})}}  H^-_{\nu_a}(-2M\omega,\omega r)  
\exp \! \left[ -i \left(-2M\omega \log (4M\omega) +  \tfrac{\pi}{2}\nu_a - \sigma_{\nu_a}(-{2M}\omega)\right) \right], 
\end{align}
\end{subequations}
\end{widetext}
where $H^\pm$ are Coulomb wave functions, the Coulomb phase shift is 
\begin{equation}
\label{cshift}
\sigma_{\nu}(\eta)=-\frac{i}{2}\log\frac{\Gamma(\nu+1+i\eta)}{\Gamma(\nu+1-i\eta)},
\end{equation}
and $\nu_a=\frac{1}{2}(-1 + \sqrt{(2 l + 1)^2 - 12 (2M\omega)^2})$. The final factor ensures that the phase 
(i.e.\ complex argument) 
of the mode matches the asymptotic solution $e^{\pm i \omega r}$. Note that some other references use a different expression for $\sigma$ involving arg; such expressions are only valid for real~$\omega$.

On the other hand, we estimate the in and out modes near the horizon using the Jaff\'{e} power series solution given in Eqs.\ (49)--(51) 
of~\cite{Leaver1986}. 
For the in modes,
\begin{equation}
\phi_{\omega l}^{in} \sim \frac{1}{r}e^{-i\omega r}
\left(\frac{r}{2M}-1\right)^{-i 2M\omega}\sum_{n=0}^{\infty}a_n \left( 1-\frac{2M}{r}\right)^n , 
\end{equation}
where
\begin{align}
a_{-1} &= 0, \qquad 
a_0 = 1, \nonumber \\
&(1 + n) (1 + n - 2 i (2M\omega)) a_{n+1} \nonumber \\
+ &(-1 - l (l + 1) - 2 n (1 + n)) a_n + n^2 a_{n - 1} = 0,
\end{align}
and for the out modes we replace $\omega$ by $-\omega$ in the equations above.
While these solutions are, in principle, analytic, in practice the convergence becomes extremely slow far from the horizon; as a result, it is more practical to use them to find the asymptotic value of the modes. Specifically, we evaluate the Jaff\'{e} asymptotic expressions at fixed  boundary values of $r$ just outside and inside the horizon,
then use the differential equation to extrapolate further away. Our results are robust to the choice of values for these two parameters.

Unfortunately, this numerical integration method is not without issues. The most serious problems occur when the energy of the mode is very small compared to the angular momentum $l$ -- or rather, the effective potential $V_l(r)$. In that case, continuing the analogy with one-dimensional scattering, the incident and reflected waves will have similar magnitudes, while the transmitted wave will be very, very small;  the relative disparity in magnitudes amplifies numerical imprecision. Some care is therefore needed.

\section{Convergence of the mode sums and integrals}
\label{app:convergence} 

In this appendix we describe the numerical accuracy in the mode sums and integrals. 

The evaluation of the response function is done mode by mode, and each `mode' makes a positive contribution to the  response function. Therefore, calculating the  response function is a simple matter of summing over the contributions of all modes, questions of convergence notwithstanding. It thus remains to show, analytically or otherwise, that the sum over $l$ and $\omega$ remains finite; while a suitably chosen switching function will accelerate convergence over~$\omega$, we must carefully consider the sum over~$l$. Outside, of course, the sum over $l$ behaves similarly to the Minkowski case, but the sum inside the black hole needs more care.

In practice, our program proceeds as follows: we evaluate the mode functions for each $l$ on a discretized $\omega$ grid, which is then interpolated and integrated over, taking care to evaluate at enough $\omega$ points to ensure smoothness of the interpolating function. (The low $\omega$ regime, in particular, requires evaluation to high resolution in order to capture the features of the sharp peaks found there.) These $\omega$ integrals are then summed over to a suitable maximum~$l$.

This approach raises a technical issue: by summing over modes in $\omega$ and $l$, we are effectively summing in two infinite dimensions. We deal with this by putting in  cut-offs on $\omega$ and $l$ such that any term  beyond the cutoff values have contributions smaller than $0.0001\%$ of the total sum. We find that in general the cut-off $l_{\rm max}$ grows with larger distances, smaller switching widths or decreasing energy gaps. We have also checked the robustness of our results  by refining the grid of the $\omega$ summation.

\section{Comparison with a static detector in Minkowski spacetime}
\label{app:minkowski-comparison} 

In this appendix we compare our infalling detector results to the response of a static detector in a thermal state in Minkowski spacetime, and, as the zero temperature limit, to the response of an inertial detector in Minkowski vacuum. 

\begin{figure}
    \centering
    \includegraphics[width=0.9\columnwidth]{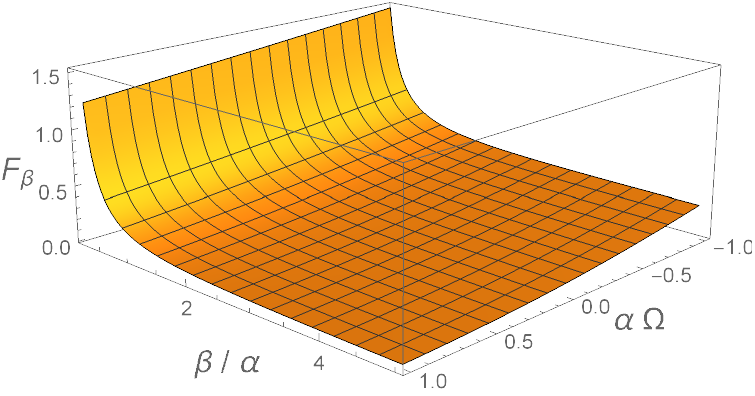}
    \caption{Plot of the Minkowski thermal state response $F_\beta$ \eqref{eq:Fbeta-minkowski} as a function of $\alpha\Omega$ and
    $\beta/\alpha$.\label{fig:Minkbetaplot}}
\end{figure}



Starting with the general expression \eqref{eq:responsefunction-formula} for the response function, and assuming that $W(\tau,\tau')$ depends on its two arguments only through their difference, $W(\tau,\tau') = W(\tau-\tau')$, the response function takes the form \cite{Fewster:2016ewy}
\begin{align}
F(\Omega) = \frac{1}{2\pi} \int_{-\infty}^\infty \textrm{d}\omega 
\left|\widehat\chi(\omega) \right|^2 \widehat{W}(\Omega-\omega), 
\end{align}
where the hat stands for the Fourier transform, in the convention 
\begin{align}
\widehat\chi(\omega) = \int_{-\infty}^\infty \textrm{d}\tau \, e^{-i\omega\tau} \chi(\tau). 
\end{align}

We now specialise to four-dimensional Minkowski spacetime. For a detector that is static in the rest frame of a thermal state of temperature $\beta^{-1}>0$, 
we have \cite{Takagi:1986kn}
\begin{align}
\widehat{W}_\beta(\omega) = \frac{\omega}{2\pi \! \left(e^{\beta\omega} -1\right)}. 
\end{align}
We write the switching function \eqref{eq:switching-family} as 
\begin{align}
\chi(\tau)=\begin{cases}
\cos^4 \bigl(\tau/(2\alpha)\bigr)&
\text{for} \ \ |\tau| < \pi\alpha, \\
0 & \textrm{otherwise}, 
\end{cases}
\end{align}
where the positive parameter $\alpha$ is related to $\Delta$ in \eqref{eq:switching-family} by $\alpha = \Delta/\pi$, and the parameter $\tau_{mid}$ in \eqref{eq:switching-family} has been dropped, by time translation invariance of the Minkowski thermal state. 
It follows that 
\begin{align}
\widehat\chi(\omega) &= \alpha H(\alpha\omega),
\end{align}
where
\begin{align}
H(z) &= \frac{3 \sin(\pi z)}{z(1-z^2)(4-z^2)}. 
\end{align}

Combining these observations, the response function takes the form 
\begin{align}
F_\beta(\Omega) = \frac{1}{4\pi^2} \int_{-\infty}^\infty 
\frac{\textrm{d}z \, z}{1 - e^{-(\beta/\alpha) z}} \, H^2(z+\alpha\Omega). 
\label{eq:Fbeta-minkowski}
\end{align}
A plot of $F_\beta$ is given in Figure~\ref{fig:Minkbetaplot}. 
In the zero temperature limit, $\beta\to\infty$, 
$F_\beta$ reduces to the Minkowski vacuum response, 
\begin{align}
F_{vac}(\Omega) = \frac{1}{4\pi^2} \int_0^\infty \textrm{d}z
\, z H^2(z+\alpha\Omega), 
\label{eq:Fvac-minkowski}
\end{align}
which can be expressed as a sum of sine and cosine integrals~\cite{NIST:DLMF}. 

We wish to compare $F_\beta$ and $F_{vac}$ with the Schwarzschild response shown in Figure~\ref{fig:vacuumgraph}, in which $\Delta = 0.3 \times 2M$ and $\Omega = 5/(2M)$. In terms of our parameters, this means $\alpha\Omega = 1.5/\pi$, and, using for $\beta$ the Schwarzschild infinity value $8\pi M$, $\beta/\alpha = 40\pi^2/3$. With these values, we have 
$F_{vac} \approx 0.0334638$ and 
$F_\beta \approx 0.0334861$, shown as horizontal dashed lines in Figure~\ref{fig:vacuumgraph}. 

In the Schwarzschild responses shown in 
Figures \ref{fig:deltagraph} and~\ref{fig:omegagraph}, 
the normalisation at infinity 
is obtained by setting $F_{\text{asymp}}$ to $F_\beta$ 
with $\beta = 8\pi M$ and $\alpha = \Delta/\pi$.

\bibliography{infall_draft}

\begin{thebibliography}{34}%
\makeatletter
\providecommand \@ifxundefined [1]{%
 \@ifx{#1\undefined}
}%
\providecommand \@ifnum [1]{%
 \ifnum #1\expandafter \@firstoftwo
 \else \expandafter \@secondoftwo
 \fi
}%
\providecommand \@ifx [1]{%
 \ifx #1\expandafter \@firstoftwo
 \else \expandafter \@secondoftwo
 \fi
}%
\providecommand \natexlab [1]{#1}%
\providecommand \enquote  [1]{``#1''}%
\providecommand \bibnamefont  [1]{#1}%
\providecommand \bibfnamefont [1]{#1}%
\providecommand \citenamefont [1]{#1}%
\providecommand \href@noop [0]{\@secondoftwo}%
\providecommand \href [0]{\begingroup \@sanitize@url \@href}%
\providecommand \@href[1]{\@@startlink{#1}\@@href}%
\providecommand \@@href[1]{\endgroup#1\@@endlink}%
\providecommand \@sanitize@url [0]{\catcode `\\12\catcode `\$12\catcode
  `\&12\catcode `\#12\catcode `\^12\catcode `\_12\catcode `\%12\relax}%
\providecommand \@@startlink[1]{}%
\providecommand \@@endlink[0]{}%
\providecommand \url  [0]{\begingroup\@sanitize@url \@url }%
\providecommand \@url [1]{\endgroup\@href {#1}{\urlprefix }}%
\providecommand \urlprefix  [0]{URL }%
\providecommand \Eprint [0]{\href }%
\providecommand \doibase [0]{https://doi.org/}%
\providecommand \selectlanguage [0]{\@gobble}%
\providecommand \bibinfo  [0]{\@secondoftwo}%
\providecommand \bibfield  [0]{\@secondoftwo}%
\providecommand \translation [1]{[#1]}%
\providecommand \BibitemOpen [0]{}%
\providecommand \bibitemStop [0]{}%
\providecommand \bibitemNoStop [0]{.\EOS\space}%
\providecommand \EOS [0]{\spacefactor3000\relax}%
\providecommand \BibitemShut  [1]{\csname bibitem#1\endcsname}%
\let\auto@bib@innerbib\@empty
\bibitem [{\citenamefont {Mathur}(2009)}]{Mathur:2009hf}%
  \BibitemOpen
  \bibfield  {author} {\bibinfo {author} {\bibfnamefont {S.~D.}\ \bibnamefont
  {Mathur}},\ }\bibfield  {title} {\bibinfo {title} {{The Information paradox:
  A Pedagogical introduction}},\ }\href
  {https://doi.org/10.1088/0264-9381/26/22/224001} {\bibfield  {journal}
  {\bibinfo  {journal} {Class. Quant. Grav.}\ }\textbf {\bibinfo {volume}
  {26}},\ \bibinfo {pages} {224001} (\bibinfo {year} {2009})},\ \Eprint
  {https://arxiv.org/abs/0909.1038} {arXiv:0909.1038 [hep-th]} \BibitemShut
  {NoStop}%
\bibitem [{\citenamefont {Mann}(2015)}]{Mann:2015luq}%
  \BibitemOpen
  \bibfield  {author} {\bibinfo {author} {\bibfnamefont {R.~B.}\ \bibnamefont
  {Mann}},\ }\href {https://doi.org/10.1007/978-3-319-14496-2} {\emph {\bibinfo
  {title} {{Black Holes: Thermodynamics, Information, and Firewalls}}}},\
  SpringerBriefs in Physics\ (\bibinfo  {publisher} {Springer},\ \bibinfo
  {year} {2015})\BibitemShut {NoStop}%
\bibitem [{\citenamefont {Almheiri}\ \emph {et~al.}(2013)\citenamefont
  {Almheiri}, \citenamefont {Marolf}, \citenamefont {Polchinski},\ and\
  \citenamefont {Sully}}]{Almheiri:2012rt}%
  \BibitemOpen
  \bibfield  {author} {\bibinfo {author} {\bibfnamefont {A.}~\bibnamefont
  {Almheiri}}, \bibinfo {author} {\bibfnamefont {D.}~\bibnamefont {Marolf}},
  \bibinfo {author} {\bibfnamefont {J.}~\bibnamefont {Polchinski}},\ and\
  \bibinfo {author} {\bibfnamefont {J.}~\bibnamefont {Sully}},\ }\bibfield
  {title} {\bibinfo {title} {{Black Holes: Complementarity or Firewalls?}},\
  }\href {https://doi.org/10.1007/JHEP02(2013)062} {\bibfield  {journal}
  {\bibinfo  {journal} {JHEP}\ }\textbf {\bibinfo {volume} {02}},\ \bibinfo
  {pages} {062}},\ \Eprint {https://arxiv.org/abs/1207.3123} {arXiv:1207.3123
  [hep-th]} \BibitemShut {NoStop}%
\bibitem [{\citenamefont {Hawking}(1975)}]{Hawking:1975vcx}%
  \BibitemOpen
  \bibfield  {author} {\bibinfo {author} {\bibfnamefont {S.~W.}\ \bibnamefont
  {Hawking}},\ }\bibfield  {title} {\bibinfo {title} {{Particle Creation by
  Black Holes}},\ }\href {https://doi.org/10.1007/BF02345020} {\bibfield
  {journal} {\bibinfo  {journal} {Commun. Math. Phys.}\ }\textbf {\bibinfo
  {volume} {43}},\ \bibinfo {pages} {199} (\bibinfo {year} {1975})},\ \bibinfo
  {note} {[Erratum: Commun.Math.Phys. 46, 206 (1976)]}\BibitemShut {NoStop}%
\bibitem [{\citenamefont {Unruh}(1976)}]{unruh1976}%
  \BibitemOpen
  \bibfield  {author} {\bibinfo {author} {\bibfnamefont {W.~G.}\ \bibnamefont
  {Unruh}},\ }\bibfield  {title} {\bibinfo {title} {Notes on black-hole
  evaporation},\ }\href {https://doi.org/10.1103/PhysRevD.14.870} {\bibfield
  {journal} {\bibinfo  {journal} {Phys. Rev. D}\ }\textbf {\bibinfo {volume}
  {14}},\ \bibinfo {pages} {870} (\bibinfo {year} {1976})}\BibitemShut
  {NoStop}%
\bibitem [{\citenamefont {Hartle}\ and\ \citenamefont
  {Hawking}(1976)}]{Hartle:1976tp}%
  \BibitemOpen
  \bibfield  {author} {\bibinfo {author} {\bibfnamefont {J.~B.}\ \bibnamefont
  {Hartle}}\ and\ \bibinfo {author} {\bibfnamefont {S.~W.}\ \bibnamefont
  {Hawking}},\ }\bibfield  {title} {\bibinfo {title} {{Path Integral Derivation
  of Black Hole Radiance}},\ }\href {https://doi.org/10.1103/PhysRevD.13.2188}
  {\bibfield  {journal} {\bibinfo  {journal} {Phys. Rev. D}\ }\textbf {\bibinfo
  {volume} {13}},\ \bibinfo {pages} {2188} (\bibinfo {year}
  {1976})}\BibitemShut {NoStop}%
\bibitem [{\citenamefont {Israel}(1976)}]{Israel:1976ur}%
  \BibitemOpen
  \bibfield  {author} {\bibinfo {author} {\bibfnamefont {W.}~\bibnamefont
  {Israel}},\ }\bibfield  {title} {\bibinfo {title} {{Thermo field dynamics of
  black holes}},\ }\href {https://doi.org/10.1016/0375-9601(76)90178-X}
  {\bibfield  {journal} {\bibinfo  {journal} {Phys. Lett. A}\ }\textbf
  {\bibinfo {volume} {57}},\ \bibinfo {pages} {107} (\bibinfo {year}
  {1976})}\BibitemShut {NoStop}%
\bibitem [{\citenamefont {Hu}\ \emph {et~al.}(2012)\citenamefont {Hu},
  \citenamefont {Lin},\ and\ \citenamefont {Louko}}]{Hu:2012jr}%
  \BibitemOpen
  \bibfield  {author} {\bibinfo {author} {\bibfnamefont {B.~L.}\ \bibnamefont
  {Hu}}, \bibinfo {author} {\bibfnamefont {S.-Y.}\ \bibnamefont {Lin}},\ and\
  \bibinfo {author} {\bibfnamefont {J.}~\bibnamefont {Louko}},\ }\bibfield
  {title} {\bibinfo {title} {{Relativistic Quantum Information in
  Detectors-Field Interactions}},\ }\href
  {https://doi.org/10.1088/0264-9381/29/22/224005} {\bibfield  {journal}
  {\bibinfo  {journal} {Class. Quant. Grav.}\ }\textbf {\bibinfo {volume}
  {29}},\ \bibinfo {pages} {224005} (\bibinfo {year} {2012})},\ \Eprint
  {https://arxiv.org/abs/1205.1328} {arXiv:1205.1328 [quant-ph]} \BibitemShut
  {NoStop}%
\bibitem [{\citenamefont {DeWitt}(1979)}]{DeWitt1979}%
  \BibitemOpen
  \bibfield  {author} {\bibinfo {author} {\bibfnamefont {B.~S.}\ \bibnamefont
  {DeWitt}},\ }\bibfield  {title} {\bibinfo {title} {Quantum gravity: the new
  synthesis},\ }in\ \href@noop {} {\emph {\bibinfo {booktitle} {General
  Relativity: an Einstein centenary survey}}},\ \bibinfo {editor} {edited by\
  \bibinfo {editor} {\bibfnamefont {S.~W.}\ \bibnamefont {Hawking}}\ and\
  \bibinfo {editor} {\bibfnamefont {W.}~\bibnamefont {Israel}}}\ (\bibinfo
  {publisher} {Cambridge University Press},\ \bibinfo {address} {Cambridge},\
  \bibinfo {year} {1979})\BibitemShut {NoStop}%
\bibitem [{\citenamefont {Hodgkinson}\ and\ \citenamefont
  {Louko}(2012)}]{Hodgkinson:2012mr}%
  \BibitemOpen
  \bibfield  {author} {\bibinfo {author} {\bibfnamefont {L.}~\bibnamefont
  {Hodgkinson}}\ and\ \bibinfo {author} {\bibfnamefont {J.}~\bibnamefont
  {Louko}},\ }\bibfield  {title} {\bibinfo {title} {{Static, stationary and
  inertial Unruh-DeWitt detectors on the BTZ black hole}},\ }\href
  {https://doi.org/10.1103/PhysRevD.86.064031} {\bibfield  {journal} {\bibinfo
  {journal} {Phys. Rev. D}\ }\textbf {\bibinfo {volume} {86}},\ \bibinfo
  {pages} {064031} (\bibinfo {year} {2012})},\ \Eprint
  {https://arxiv.org/abs/1206.2055} {arXiv:1206.2055 [gr-qc]} \BibitemShut
  {NoStop}%
\bibitem [{\citenamefont {Hodgkinson}\ \emph {et~al.}(2014)\citenamefont
  {Hodgkinson}, \citenamefont {Louko},\ and\ \citenamefont
  {Ottewill}}]{Hodgkinson:2014iua}%
  \BibitemOpen
  \bibfield  {author} {\bibinfo {author} {\bibfnamefont {L.}~\bibnamefont
  {Hodgkinson}}, \bibinfo {author} {\bibfnamefont {J.}~\bibnamefont {Louko}},\
  and\ \bibinfo {author} {\bibfnamefont {A.~C.}\ \bibnamefont {Ottewill}},\
  }\bibfield  {title} {\bibinfo {title} {{Static detectors and
  circular-geodesic detectors on the Schwarzschild black hole}},\ }\href
  {https://doi.org/10.1103/PhysRevD.89.104002} {\bibfield  {journal} {\bibinfo
  {journal} {Phys. Rev. D}\ }\textbf {\bibinfo {volume} {89}},\ \bibinfo
  {pages} {104002} (\bibinfo {year} {2014})},\ \Eprint
  {https://arxiv.org/abs/1401.2667} {arXiv:1401.2667 [gr-qc]} \BibitemShut
  {NoStop}%
\bibitem [{\citenamefont {Ju\'arez-Aubry}\ and\ \citenamefont
  {Louko}(2014)}]{Juarez-Aubry:2014jba}%
  \BibitemOpen
  \bibfield  {author} {\bibinfo {author} {\bibfnamefont {B.~A.}\ \bibnamefont
  {Ju\'arez-Aubry}}\ and\ \bibinfo {author} {\bibfnamefont {J.}~\bibnamefont
  {Louko}},\ }\bibfield  {title} {\bibinfo {title} {{Onset and decay of the 1 +
  1 Hawking-Unruh effect: what the derivative-coupling detector saw}},\ }\href
  {https://doi.org/10.1088/0264-9381/31/24/245007} {\bibfield  {journal}
  {\bibinfo  {journal} {Class. Quant. Grav.}\ }\textbf {\bibinfo {volume}
  {31}},\ \bibinfo {pages} {245007} (\bibinfo {year} {2014})},\ \Eprint
  {https://arxiv.org/abs/1406.2574} {arXiv:1406.2574 [gr-qc]} \BibitemShut
  {NoStop}%
\bibitem [{\citenamefont {Ju\'arez-Aubry}(2017)}]{JuarezPhDThesis}%
  \BibitemOpen
  \bibfield  {author} {\bibinfo {author} {\bibfnamefont {B.~A.}\ \bibnamefont
  {Ju\'arez-Aubry}},\ }\emph {\bibinfo {title} {{Asymptotics in the
  time-dependent Hawking and Unruh effects}}},\ \href@noop {} {Ph.D. thesis},\
  \bibinfo  {school} {University of Nottingham} (\bibinfo {year} {2017}),\
  \Eprint {https://arxiv.org/abs/1708.09430} {arXiv:1708.09430 [gr-qc]}
  \BibitemShut {NoStop}%
\bibitem [{\citenamefont {Ju\'arez-Aubry}\ and\ \citenamefont
  {Louko}(2022)}]{Juarez-Aubry:2021tae}%
  \BibitemOpen
  \bibfield  {author} {\bibinfo {author} {\bibfnamefont {B.~A.}\ \bibnamefont
  {Ju\'arez-Aubry}}\ and\ \bibinfo {author} {\bibfnamefont {J.}~\bibnamefont
  {Louko}},\ }\bibfield  {title} {\bibinfo {title} {{Quantum kicks near a
  Cauchy horizon}},\ }\href {https://doi.org/10.1116/5.0073373} {\bibfield
  {journal} {\bibinfo  {journal} {AVS Quantum Sci.}\ }\textbf {\bibinfo
  {volume} {4}},\ \bibinfo {pages} {013201} (\bibinfo {year} {2022})},\ \Eprint
  {https://arxiv.org/abs/2109.14601} {arXiv:2109.14601 [gr-qc]} \BibitemShut
  {NoStop}%
\bibitem [{\citenamefont {Gallock-Yoshimura}\ \emph {et~al.}(2021)\citenamefont
  {Gallock-Yoshimura}, \citenamefont {Tjoa},\ and\ \citenamefont
  {Mann}}]{Gallock-Yoshimura:2021yok}%
  \BibitemOpen
  \bibfield  {author} {\bibinfo {author} {\bibfnamefont {K.}~\bibnamefont
  {Gallock-Yoshimura}}, \bibinfo {author} {\bibfnamefont {E.}~\bibnamefont
  {Tjoa}},\ and\ \bibinfo {author} {\bibfnamefont {R.~B.}\ \bibnamefont
  {Mann}},\ }\bibfield  {title} {\bibinfo {title} {{Harvesting entanglement
  with detectors freely falling into a black hole}},\ }\href
  {https://doi.org/10.1103/PhysRevD.104.025001} {\bibfield  {journal} {\bibinfo
   {journal} {Phys. Rev. D}\ }\textbf {\bibinfo {volume} {104}},\ \bibinfo
  {pages} {025001} (\bibinfo {year} {2021})},\ \Eprint
  {https://arxiv.org/abs/2102.09573} {arXiv:2102.09573 [quant-ph]} \BibitemShut
  {NoStop}%
\bibitem [{\citenamefont {Boulware}(1975)}]{Boulware:1974dm}%
  \BibitemOpen
  \bibfield  {author} {\bibinfo {author} {\bibfnamefont {D.~G.}\ \bibnamefont
  {Boulware}},\ }\bibfield  {title} {\bibinfo {title} {{Quantum Field Theory in
  Schwarzschild and Rindler Spaces}},\ }\href
  {https://doi.org/10.1103/PhysRevD.11.1404} {\bibfield  {journal} {\bibinfo
  {journal} {Phys. Rev. D}\ }\textbf {\bibinfo {volume} {11}},\ \bibinfo
  {pages} {1404} (\bibinfo {year} {1975})}\BibitemShut {NoStop}%
\bibitem [{\citenamefont {Birrell}\ and\ \citenamefont
  {Davies}(1982)}]{birrell-davies1982}%
  \BibitemOpen
  \bibfield  {author} {\bibinfo {author} {\bibfnamefont {N.~D.}\ \bibnamefont
  {Birrell}}\ and\ \bibinfo {author} {\bibfnamefont {P.~C.~W.}\ \bibnamefont
  {Davies}},\ }\href@noop {} {\emph {\bibinfo {title} {Quantum Fields in Curved
  Space}}}\ (\bibinfo  {publisher} {Cambridge University Press},\ \bibinfo
  {address} {Cambridge},\ \bibinfo {year} {1982})\BibitemShut {NoStop}%
\bibitem [{\citenamefont {Mart\'\i{}n-Mart\'\i{}nez}\ \emph
  {et~al.}(2013)\citenamefont {Mart\'\i{}n-Mart\'\i{}nez}, \citenamefont
  {Montero},\ and\ \citenamefont {del Rey}}]{MartinMartinez:2012th}%
  \BibitemOpen
  \bibfield  {author} {\bibinfo {author} {\bibfnamefont {E.}~\bibnamefont
  {Mart\'\i{}n-Mart\'\i{}nez}}, \bibinfo {author} {\bibfnamefont
  {M.}~\bibnamefont {Montero}},\ and\ \bibinfo {author} {\bibfnamefont
  {M.}~\bibnamefont {del Rey}},\ }\bibfield  {title} {\bibinfo {title}
  {{Wavepacket detection with the Unruh-DeWitt model}},\ }\href
  {https://doi.org/10.1103/PhysRevD.87.064038} {\bibfield  {journal} {\bibinfo
  {journal} {Phys. Rev. D}\ }\textbf {\bibinfo {volume} {87}},\ \bibinfo
  {pages} {064038} (\bibinfo {year} {2013})},\ \Eprint
  {https://arxiv.org/abs/1207.3248} {arXiv:1207.3248 [quant-ph]} \BibitemShut
  {NoStop}%
\bibitem [{\citenamefont {Alhambra}\ \emph {et~al.}(2014)\citenamefont
  {Alhambra}, \citenamefont {Kempf},\ and\ \citenamefont
  {Mart\'\i{}n-Mart\'\i{}nez}}]{Alhambra:2013uja}%
  \BibitemOpen
  \bibfield  {author} {\bibinfo {author} {\bibfnamefont {{\'A}.~M.}\
  \bibnamefont {Alhambra}}, \bibinfo {author} {\bibfnamefont {A.}~\bibnamefont
  {Kempf}},\ and\ \bibinfo {author} {\bibfnamefont {E.}~\bibnamefont
  {Mart\'\i{}n-Mart\'\i{}nez}},\ }\bibfield  {title} {\bibinfo {title}
  {{Casimir forces on atoms in optical cavities}},\ }\href
  {https://doi.org/10.1103/PhysRevA.89.033835} {\bibfield  {journal} {\bibinfo
  {journal} {Phys. Rev. A}\ }\textbf {\bibinfo {volume} {89}},\ \bibinfo
  {pages} {033835} (\bibinfo {year} {2014})},\ \Eprint
  {https://arxiv.org/abs/1311.7619} {arXiv:1311.7619 [quant-ph]} \BibitemShut
  {NoStop}%
\bibitem [{\citenamefont {{D\'ecanini}}\ and\ \citenamefont
  {Folacci}(2008)}]{Decanini:2005eg}%
  \BibitemOpen
  \bibfield  {author} {\bibinfo {author} {\bibfnamefont {Y.}~\bibnamefont
  {{D\'ecanini}}}\ and\ \bibinfo {author} {\bibfnamefont {A.}~\bibnamefont
  {Folacci}},\ }\bibfield  {title} {\bibinfo {title} {{Hadamard renormalization
  of the stress-energy tensor for a quantized scalar field in a general
  spacetime of arbitrary dimension}},\ }\href
  {https://doi.org/10.1103/PhysRevD.78.044025} {\bibfield  {journal} {\bibinfo
  {journal} {Phys. Rev. D}\ }\textbf {\bibinfo {volume} {78}},\ \bibinfo
  {pages} {044025} (\bibinfo {year} {2008})},\ \Eprint
  {https://arxiv.org/abs/gr-qc/0512118} {arXiv:gr-qc/0512118} \BibitemShut
  {NoStop}%
\bibitem [{\citenamefont {H{\"o}rmander}(1986)}]{hormander-book}%
  \BibitemOpen
  \bibfield  {author} {\bibinfo {author} {\bibfnamefont {L.}~\bibnamefont
  {H{\"o}rmander}},\ }\href@noop {} {\emph {\bibinfo {title} {The Analysis of
  Linear Partial Differential Operators}}}\ (\bibinfo  {publisher} {Springer},\
  \bibinfo {address} {Berlin},\ \bibinfo {year} {1986})\BibitemShut {NoStop}%
\bibitem [{\citenamefont {Fewster}(2000)}]{Fewster:1999gj}%
  \BibitemOpen
  \bibfield  {author} {\bibinfo {author} {\bibfnamefont {C.~J.}\ \bibnamefont
  {Fewster}},\ }\bibfield  {title} {\bibinfo {title} {{A General worldline
  quantum inequality}},\ }\href {https://doi.org/10.1088/0264-9381/17/9/302}
  {\bibfield  {journal} {\bibinfo  {journal} {Class. Quant. Grav.}\ }\textbf
  {\bibinfo {volume} {17}},\ \bibinfo {pages} {1897} (\bibinfo {year}
  {2000})},\ \Eprint {https://arxiv.org/abs/gr-qc/9910060}
  {arXiv:gr-qc/9910060} \BibitemShut {NoStop}%
\bibitem [{\citenamefont {Leaver}(1986)}]{Leaver1986}%
  \BibitemOpen
  \bibfield  {author} {\bibinfo {author} {\bibfnamefont {E.~W.}\ \bibnamefont
  {Leaver}},\ }\bibfield  {title} {\bibinfo {title} {Solutions to a generalized
  spheroidal wave equation: Teukolsky’s equations in general relativity, and
  the two‐center problem in molecular quantum mechanics},\ }\href
  {https://doi.org/10.1063/1.527130} {\bibfield  {journal} {\bibinfo  {journal}
  {J. Math. Phys.}\ }\textbf {\bibinfo {volume} {27}},\ \bibinfo {pages} {1238}
  (\bibinfo {year} {1986})}\BibitemShut {NoStop}%
\bibitem [{\citenamefont {Casals}\ \emph {et~al.}(2013)\citenamefont {Casals},
  \citenamefont {Dolan}, \citenamefont {Nolan}, \citenamefont {Ottewill},\ and\
  \citenamefont {Winstanley}}]{Casals:2012es}%
  \BibitemOpen
  \bibfield  {author} {\bibinfo {author} {\bibfnamefont {M.}~\bibnamefont
  {Casals}}, \bibinfo {author} {\bibfnamefont {S.~R.}\ \bibnamefont {Dolan}},
  \bibinfo {author} {\bibfnamefont {B.~C.}\ \bibnamefont {Nolan}}, \bibinfo
  {author} {\bibfnamefont {A.~C.}\ \bibnamefont {Ottewill}},\ and\ \bibinfo
  {author} {\bibfnamefont {E.}~\bibnamefont {Winstanley}},\ }\bibfield  {title}
  {\bibinfo {title} {{Quantization of fermions on Kerr space-time}},\ }\href
  {https://doi.org/10.1103/PhysRevD.87.064027} {\bibfield  {journal} {\bibinfo
  {journal} {Phys. Rev. D}\ }\textbf {\bibinfo {volume} {87}},\ \bibinfo
  {pages} {064027} (\bibinfo {year} {2013})},\ \Eprint
  {https://arxiv.org/abs/1207.7089} {arXiv:1207.7089 [gr-qc]} \BibitemShut
  {NoStop}%
\bibitem [{\citenamefont {Hodgkinson}(2013)}]{HodgkinsonPhDThesis}%
  \BibitemOpen
  \bibfield  {author} {\bibinfo {author} {\bibfnamefont {L.}~\bibnamefont
  {Hodgkinson}},\ }\emph {\bibinfo {title} {Particle detectors in curved
  spacetime quantum field theory}},\ \href@noop {} {Ph.D. thesis},\ \bibinfo
  {school} {University of Nottingham} (\bibinfo {year} {2013}),\ \Eprint
  {https://arxiv.org/abs/1309.7281 [gr-qc]} {arXiv:1309.7281 [gr-qc]}
  \BibitemShut {NoStop}%
\bibitem [{\citenamefont {Lanir}\ \emph {et~al.}(2018)\citenamefont {Lanir},
  \citenamefont {Levi},\ and\ \citenamefont {Ori}}]{Lanir:2018rap}%
  \BibitemOpen
  \bibfield  {author} {\bibinfo {author} {\bibfnamefont {A.}~\bibnamefont
  {Lanir}}, \bibinfo {author} {\bibfnamefont {A.}~\bibnamefont {Levi}},\ and\
  \bibinfo {author} {\bibfnamefont {A.}~\bibnamefont {Ori}},\ }\bibfield
  {title} {\bibinfo {title} {{Mode-sum renormalization of
  $\langle\Phi^{2}\rangle$ for a quantum scalar field inside a Schwarzschild
  black hole}},\ }\href {https://doi.org/10.1103/PhysRevD.98.084017} {\bibfield
   {journal} {\bibinfo  {journal} {Phys. Rev. D}\ }\textbf {\bibinfo {volume}
  {98}},\ \bibinfo {pages} {084017} (\bibinfo {year} {2018})},\ \Eprint
  {https://arxiv.org/abs/1808.06195} {arXiv:1808.06195 [gr-qc]} \BibitemShut
  {NoStop}%
\bibitem [{\citenamefont {Cong}\ \emph {et~al.}(2020)\citenamefont {Cong},
  \citenamefont {Bi\v{c}ak}, \citenamefont {Kubiz\v{n}\'ak},\ and\
  \citenamefont {Mann}}]{Cong:2020crf}%
  \BibitemOpen
  \bibfield  {author} {\bibinfo {author} {\bibfnamefont {W.}~\bibnamefont
  {Cong}}, \bibinfo {author} {\bibfnamefont {J.}~\bibnamefont {Bi\v{c}ak}},
  \bibinfo {author} {\bibfnamefont {D.}~\bibnamefont {Kubiz\v{n}\'ak}},\ and\
  \bibinfo {author} {\bibfnamefont {R.~B.}\ \bibnamefont {Mann}},\ }\bibfield
  {title} {\bibinfo {title} {{Quantum distinction of inertial frames: Local
  versus global}},\ }\href {https://doi.org/10.1103/PhysRevD.101.104060}
  {\bibfield  {journal} {\bibinfo  {journal} {Phys. Rev. D}\ }\textbf {\bibinfo
  {volume} {101}},\ \bibinfo {pages} {104060} (\bibinfo {year} {2020})},\
  \Eprint {https://arxiv.org/abs/2003.09719} {arXiv:2003.09719 [gr-qc]}
  \BibitemShut {NoStop}%
\bibitem [{\citenamefont {Mart\'{i}n-Mart\'{i}nez}\ and\ \citenamefont
  {Louko}(2014)}]{Martin-Martinez:2014qda}%
  \BibitemOpen
  \bibfield  {author} {\bibinfo {author} {\bibfnamefont {E.}~\bibnamefont
  {Mart\'{i}n-Mart\'{i}nez}}\ and\ \bibinfo {author} {\bibfnamefont
  {J.}~\bibnamefont {Louko}},\ }\bibfield  {title} {\bibinfo {title} {{Particle
  detectors and the zero mode of a quantum field}},\ }\href
  {https://doi.org/10.1103/PhysRevD.90.024015} {\bibfield  {journal} {\bibinfo
  {journal} {Phys. Rev. D}\ }\textbf {\bibinfo {volume} {90}},\ \bibinfo
  {pages} {024015} (\bibinfo {year} {2014})},\ \Eprint
  {https://arxiv.org/abs/1404.5621} {arXiv:1404.5621 [quant-ph]} \BibitemShut
  {NoStop}%
\bibitem [{\citenamefont {Fewster}\ \emph {et~al.}(2016)\citenamefont
  {Fewster}, \citenamefont {Ju\'arez-Aubry},\ and\ \citenamefont
  {Louko}}]{Fewster:2016ewy}%
  \BibitemOpen
  \bibfield  {author} {\bibinfo {author} {\bibfnamefont {C.~J.}\ \bibnamefont
  {Fewster}}, \bibinfo {author} {\bibfnamefont {B.~A.}\ \bibnamefont
  {Ju\'arez-Aubry}},\ and\ \bibinfo {author} {\bibfnamefont {J.}~\bibnamefont
  {Louko}},\ }\bibfield  {title} {\bibinfo {title} {{Waiting for Unruh}},\
  }\href {https://doi.org/10.1088/0264-9381/33/16/165003} {\bibfield  {journal}
  {\bibinfo  {journal} {Class. Quant. Grav.}\ }\textbf {\bibinfo {volume}
  {33}},\ \bibinfo {pages} {165003} (\bibinfo {year} {2016})},\ \Eprint
  {https://arxiv.org/abs/1605.01316} {arXiv:1605.01316 [gr-qc]} \BibitemShut
  {NoStop}%
\bibitem [{\citenamefont {Barbado}\ \emph {et~al.}(2011)\citenamefont
  {Barbado}, \citenamefont {Barcel\'o},\ and\ \citenamefont
  {Garay}}]{Barbado:2011dx}%
  \BibitemOpen
  \bibfield  {author} {\bibinfo {author} {\bibfnamefont {L.~C.}\ \bibnamefont
  {Barbado}}, \bibinfo {author} {\bibfnamefont {C.}~\bibnamefont {Barcel\'o}},\
  and\ \bibinfo {author} {\bibfnamefont {L.~J.}\ \bibnamefont {Garay}},\
  }\bibfield  {title} {\bibinfo {title} {{Hawking radiation as perceived by
  different observers}},\ }\href
  {https://doi.org/10.1088/0264-9381/28/12/125021} {\bibfield  {journal}
  {\bibinfo  {journal} {Class. Quant. Grav.}\ }\textbf {\bibinfo {volume}
  {28}},\ \bibinfo {pages} {125021} (\bibinfo {year} {2011})},\ \Eprint
  {https://arxiv.org/abs/1101.4382} {arXiv:1101.4382 [gr-qc]} \BibitemShut
  {NoStop}%
\bibitem [{\citenamefont {Ben-Benjamin}\ \emph {et~al.}(2019)\citenamefont
  {Ben-Benjamin} \emph {et~al.}}]{Ben-Benjamin:2019opz}%
  \BibitemOpen
  \bibfield  {author} {\bibinfo {author} {\bibfnamefont {J.~S.}\ \bibnamefont
  {Ben-Benjamin}} \emph {et~al.},\ }\bibfield  {title} {\bibinfo {title}
  {{Unruh Acceleration Radiation Revisited}},\ }\href
  {https://doi.org/10.1142/S0217751X19410057} {\bibfield  {journal} {\bibinfo
  {journal} {Int. J. Mod. Phys. A}\ }\textbf {\bibinfo {volume} {34}},\
  \bibinfo {pages} {1941005} (\bibinfo {year} {2019})},\ \Eprint
  {https://arxiv.org/abs/1906.01729} {arXiv:1906.01729 [quant-ph]} \BibitemShut
  {NoStop}%
\bibitem [{\citenamefont {Holdom}\ \emph {et~al.}(2021)\citenamefont {Holdom},
  \citenamefont {Mann},\ and\ \citenamefont {Zhang}}]{Holdom:2020uhf}%
  \BibitemOpen
  \bibfield  {author} {\bibinfo {author} {\bibfnamefont {B.}~\bibnamefont
  {Holdom}}, \bibinfo {author} {\bibfnamefont {R.~B.}\ \bibnamefont {Mann}},\
  and\ \bibinfo {author} {\bibfnamefont {C.}~\bibnamefont {Zhang}},\ }\bibfield
   {title} {\bibinfo {title} {{Unruh-DeWitt detector differentiation of black
  holes and exotic compact objects}},\ }\href
  {https://doi.org/10.1103/PhysRevD.103.124046} {\bibfield  {journal} {\bibinfo
   {journal} {Phys. Rev. D}\ }\textbf {\bibinfo {volume} {103}},\ \bibinfo
  {pages} {124046} (\bibinfo {year} {2021})},\ \Eprint
  {https://arxiv.org/abs/2011.10179} {arXiv:2011.10179 [gr-qc]} \BibitemShut
  {NoStop}%
\bibitem [{\citenamefont {Takagi}(1986)}]{Takagi:1986kn}%
  \BibitemOpen
  \bibfield  {author} {\bibinfo {author} {\bibfnamefont {S.}~\bibnamefont
  {Takagi}},\ }\bibfield  {title} {\bibinfo {title} {{Vacuum Noise and Stress
  Induced by Uniform Acceleration: Hawking-Unruh Effect in Rindler Manifold of
  Arbitrary Dimension}},\ }\href {https://doi.org/10.1143/PTP.88.1} {\bibfield
  {journal} {\bibinfo  {journal} {Prog. Theor. Phys. Suppl.}\ }\textbf
  {\bibinfo {volume} {88}},\ \bibinfo {pages} {1} (\bibinfo {year}
  {1986})}\BibitemShut {NoStop}%
\bibitem [{NIS()}]{NIST:DLMF}%
  \BibitemOpen
  \href {http://dlmf.nist.gov/} {\bibinfo {title} {{\it NIST Digital Library of
  Mathematical Functions}}},\ \bibinfo {howpublished} {http://dlmf.nist.gov/,
  Release 1.1.2 of 2021-06-15},\ \bibinfo {note} {{F}.~W.~J. Olver, A.~B. {Olde
  Daalhuis}, D.~W. Lozier, B.~I. Schneider, R.~F. Boisvert, C.~W. Clark, B.~R.
  Miller, B.~V. Saunders, H.~S. Cohl, and M.~A. McClain, eds.}\BibitemShut
  {Stop}%
\end{thebibliography}%


\end{document}